\documentclass[aps,prb,twocolumn,superscriptaddress,nofootinbib,longbibliography]{revtex4-2}
\usepackage{amssymb}
\usepackage{amsmath}
\usepackage{mathrsfs}
\usepackage{appendix}
\usepackage{graphicx}
\usepackage{float}
\usepackage{placeins}
\usepackage[normalem]{ulem}
\usepackage{color}
\usepackage{bm}
\usepackage[colorlinks=True,citecolor=blue,linkcolor=blue,allcolors=blue]{hyperref}

\begin{document}
	
\title{Topological metal-insulator transitions in one-dimensional non-Hermitian quasicrystals: beyond \texorpdfstring{${\cal PT}$}{PT}-symmetry}

\author{Guangjie Zhang}
\thanks{These authors contributed equally to this work.}
\affiliation{College of Physics and Optoelectronic Engineering, Ocean University of China, Qingdao 266100, China}

\author{Bing Shao}
\thanks{These authors contributed equally to this work.}
\affiliation{College of Physics and Optoelectronic Engineering, Ocean University of China, Qingdao 266100, China}

\author{Longwen Zhou}
\email{zhoulw13@u.nus.edu}
\affiliation{College of Physics and Optoelectronic Engineering, Ocean University of China, Qingdao 266100, China}
\affiliation{Qingdao Key Laboratory of Advanced Optoelectronics, Qingdao 266100, China}
\affiliation{Engineering Research Center of Advanced Marine Physical Instruments and Equipment of MOE, Qingdao 266100, China}

\author{Jiangbin Gong}
\email{phygj@nus.edu.sg}
\affiliation{Department of Physics, National University of Singapore, 117551, Singapore}
\affiliation{Centre for Quantum Technologies, National University of Singapore, 117543, Singapore}
\affiliation{MajuLab, CNRS-UCA-SU-NUS-NTU International Joint Research Unit, Singapore.}

\author{Weiwei Zhu}
\email{phyzhuw@tongji.edu.cn}
\affiliation{Center for Phononics and Thermal Energy Science, China-EU Joint Lab on Nanophononics, Shanghai Key Laboratory of Special Artificial Microstructure Materials and Technology, School of Physics Science and Engineering, Tongji University, Shanghai 200092, China}
	
\begin{abstract}
One-dimensional non-Hermitian quasicrystals with parity and time-reversal (${\cal PT}$) symmetry can simultaneously exhibit localization-delocalization transition, topological phase transition, and ${\cal PT}$-symmetry-breaking transition. This motivates us to investigate how the absence of ${\cal PT}$ symmetry impacts topological metal-insulator transitions in non-Hermitian quasicrystals. We propose a non-Hermitian quasiperiodic model that generally does not preserve ${\cal PT}$ symmetry but, in most parameter regions, hosts a generic triple-transition scenario encompassing localization, topological, and degeneracy-breaking transitions. These transitions can be understood within the dual momentum-space representation featuring asymmetric couplings. Specifically, the localization boundary is obtained from the dual momentum-space representation by comparing the Lyapunov exponent, or inverse localization length, of the reciprocal Aubry-Andr\'e model with the nonreciprocal hopping-asymmetry ratio of the dual lattice. The accompanying topological transition is characterized by a point-gap winding number defined with respect to a reference energy inside the complex spectral gap, whereas the degeneracy-breaking transition is understood from the loss of the reciprocal \(n\leftrightarrow -n\) spectral symmetry. The system also exhibits a localization-delocalization transition analogous to that in the Hermitian case, which occurs without accompanying topological or degeneracy-breaking transitions. Our work extends the topological metal-insulator transitions previously studied in ${\cal PT}$-symmetric systems to a more general class of non-Hermitian setting, and further reveals that non-Hermitian systems can host distinct types of localization behavior.
\end{abstract}
	
\maketitle
\section{Introduction}\label{sec:Int}

Quasiperiodic systems, which are aperiodic but possess long-range order, offer a clean and controllable platform for investigating metal-insulator transitions~\cite{PhysRevLett.51.1198,PhysRevLett.61.2144,PhysRevLett.120.160404} and exploring a wide variety of topological phases~\cite{PhysRevLett.109.106402,PhysRevLett.109.116404,PhysRevLett.108.220401} in one dimension, where arbitrarily weak uncorrelated disorder leads to Anderson localization~\cite{PhysRevLett.42.673} and only a limited set of pristine topological phases can exist~\cite{PhysRevB.82.115120}.
This is largely because quasiperiodic systems can often be obtained through dimensional reduction from higher-dimensional periodic structures. A paradigmatic example is the one-dimensional (1D) Aubry-Andr\'e-Harper (AAH) model~\cite{aubry1980analyticity}, which emerges from dimensional reduction of the Hofstadter model~\cite{PhysRevB.14.2239}. Inheriting features from its two-dimensional parent, the AAH model and its generalizations exhibit rich metal-insulator transitions and topological phases. In addition, quasiperiodic disorder has been utilized to realize topological Anderson insulators, which support gapped topological edge states that coexist with critical or extended bulk states~\cite{PhysRevA.105.063327,ZhanpengLu2025024204,PhysRevResearch.6.L042038}. This behavior contrasts with the conventional disorder-induced topological Anderson insulator~\cite{PhysRevLett.113.046802,science.aat3406,j9xb-vfrh}, which typically features gapless topological edge states and a fully localized bulk.

Non-Hermitian systems exhibit richer localization and topological phenomena than their Hermitian counterparts~\cite{Ashida02072020,RevModPhys.93.015005,e25101401,lin2023topological} even in one spatial dimension. Under open boundary conditions, bulk eigenstates can become edge-localized due to non-Hermitian skin effects (NHSEs)~\cite{PhysRevLett.121.086803,NHSE2,NHSE3,NHSE4}, in addition to conventional Anderson localization. Moreover, non-Hermitian systems can host metal-insulator transitions in one dimension under periodic boundary conditions \cite{NHTAI01,NHTAI02,NHTAI03,NHTAI04}, as exemplified by the disordered Hatano-Nelson model~\cite{PhysRevLett.77.570,HNM2,HNM3}. Non-Hermitian systems also support a broader range of topological phases than Hermitian systems, including both point-gapped and line-gapped topological states. According to the Bernard-LeClair symmetry classification~\cite{PhysRevX.9.041015,PhysRevB.99.235112,PhysRevB.100.144106}, non-Hermitian systems fall into $38$ distinct symmetry classes, revealing a richer topological landscape than Hermitian systems.

\begin{figure}
	\includegraphics[width=\linewidth]{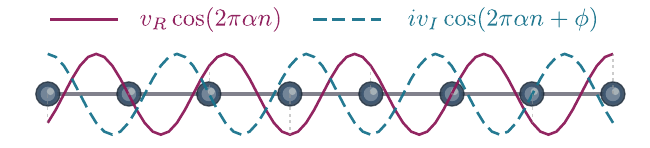}
	\caption{Real and imaginary parts of onsite potential in a one-dimensional non-Hermitian quasicrystal, both following a cosine modulation but with a relative phase shift $\phi$.}
	\label{model}
\end{figure}

The combination of quasiperiodic and non-Hermitian modulations is expected to generate unique localization and topological phenomena that are absent in either Hermitian quasicrystals or non-Hermitian crystals. Various types of such non-Hermitian quasicrystals (NHQCs) have been investigated \cite{NHTQC01,NHTQC02,NHTQC03,NHTQC04,NHTQC05,NHTQC06,NHTQC07,NHTQC08,NHTQC09,NHTQC10,NHTQC11,NHTQC12,NHTQC13,NHTQC14,NHTQC15,NHTQC16,GandhiPRB2023,NHTQC17,NHTQC18,NHTQC19,NHTQC20,8txw-m2cp}, revealing a wealth of intriguing behavior, including ${\cal PT}$-symmetry-breaking transitions, topological phase transitions, and non-Hermitian mobility edges \cite{ME01,ME02,ME03,ME04,ME05,ME06,ME07,ME08,ME09,ME10,ME11,ME12,ME13,ME14,ME15,ME16,ME17,ME18}. Among these, the topological triple phase transition \cite{TriplePT1,TriplePT2,TriplePT3} has attracted particular attention due to the simultaneous occurrence of localization, topological, and ${\cal PT}$-symmetry-breaking transitions. However, most existing studies are restricted to ${\cal PT}$-symmetric systems, and the behavior in more general non-Hermitian settings remains largely unexplored.

In contrast to previous studies on ${\cal PT}$-symmetric non-Hermitian quasicrystals, where triple phase transitions involve ${\cal PT}$-symmetry breaking, we introduce a non-Hermitian quasiperiodic model that generally lacks ${\cal PT}$ symmetry and show that it supports a generic triple transition—encompassing localization, topology, and degeneracy breaking—thereby extending the paradigm beyond the ${\cal PT}$-symmetric realm. In our model, the real and imaginary parts of the onsite potential are subject to the same cosine modulation with a relative phase shift $\phi$ (Fig.~\ref{model}). In general, such a model does not preserve the ${\cal PT}$ symmetry, except at special values $\phi = m\pi + \pi/2$ with $m \in \mathbb{Z}$. We demonstrate that our model can simultaneously exhibit localization and topological transitions by tuning either the modulation amplitude or the relative phase shift, a phenomenon that we term the topological metal-insulator transition. Furthermore, we show that the system undergoes a degeneracy-breaking transition—from twofold degenerate to non-degenerate in its spectrum—which is more general than the ${\cal PT}$ transition. All these phase transitions can be understood within the dual momentum-space representation that inherently features asymmetric couplings. The localization boundary is obtained analytically from this representation by comparing the Lyapunov exponent of the reciprocal AAH model with the hopping asymmetry of the dual lattice. An independent derivation based on the real-space Lyapunov exponent and Avila's global theory is provided in Appendix~\ref{app:avila}. The topological transition is characterized by a point-gap winding number defined with respect to a reference energy inside the complex spectral gap, whereas the degeneracy-breaking transition is understood from the loss of the reciprocal $n\leftrightarrow -n$ spectral symmetry, again in the dual space. These three transitions occur along the same boundary in most of the parameter space.

We also investigate phase transitions at special phase shifts $\phi = m\pi + \pi/2$, where the system preserves ${\cal PT}$ symmetry, and at $\phi = m\pi$, where asymmetric couplings in momentum space vanish. At $\phi = m\pi + \pi/2$, the system exhibits a quartet phase transition that involves localization, topology, degeneracy-breaking and ${\cal PT}$-breaking. At $\phi = m\pi$, only the localization transition survives. Therefore, we identify two distinct types of localization transitions in 1D NHQCs: one accompanied by point-gap topology and degeneracy breaking associated with non-Hermiticity, and the other analogous to that in Hermitian systems.

The rest of the paper is organized as follows. In Sec.~\ref{sec:Mod}, we introduce our model and the quantities to characterize its localization properties, including the inverse participation ratio (IPR) and the fractal dimension (FD). In Sec.~\ref{sec:Phase}, we examine the generic triple phase transition of localization, topology, and degeneracy-breaking by investigating IPRs, winding numbers, and degeneracy rates in the parameter space of the system. For each of the three phase transitions, we first present the numerical results and then offer a physical interpretation based on the dual momentum-space representation. In Sec.~\ref{sec:Special}, we study the localization properties of the system at specific values $\phi = m\pi + \pi/2$ and $\phi = m\pi$ of the phase shift. We conclude our study and discuss potential future work in Sec.~\ref{sec:Sum}.
Some additional theoretical and numerical details are presented in Appendices \ref{app:avila}--\ref{app:degeneracy_threshold}.

\section{Model and Methods}\label{sec:Mod}
	
We consider a 1D lattice model as illustrated in Fig.~\ref{model}, which is described by the tight-binding Hamiltonian
\begin{equation}
	H = \sum_{n} \left(t  a_{n+1}^{\dagger} a_{n} + \mathrm{H.c.} \right) + \sum_{n} V_n \, a_{n}^{\dagger} a_{n},
	\label{eq:hamiltonian}
\end{equation}
where \( a_{n} \) (\( a_{n}^{\dagger} \)) denotes the annihilation (creation) operator at the lattice site \( n \), \( t \) is the nearest-neighbor hopping amplitude, and \( V_n \) represents a complex quasiperiodic onsite potential of the form
\begin{equation}
	V_n = v_R \cos(2\pi \alpha n) + i \, v_I \cos( 2\pi \alpha n + \phi ).
	\label{eq:potential}
\end{equation}
Here, \( \alpha \) is chosen as the irrational number \((\sqrt{5}-1)/2\) to ensure quasiperiodicity, while \( v_R \), \( v_I \), and \( \phi \) denote the real amplitude, imaginary amplitude, and relative phase shift of the modulations, respectively. In general, the model does not preserve ${\cal PT}$ symmetry except at special parameter values, e.g., $\phi = m\pi + \pi/2$ with $m \in \mathbb{Z}$. For generic \(\phi\neq m\pi+\pi/2\), the Hamiltonian has no ${\cal PT}$ symmetry. In the absence of additional time-reversal, particle-hole, chiral, or pseudo-Hermitian constraints, the generic model belongs to the non-Hermitian Bernard-LeClair symmetry class A with point-gap topology \cite{PhysRevX.9.041015}. The ${\cal PT}$-symmetric points \(\phi=m\pi+\pi/2\) are symmetry-enriched special cases and will be discussed separately in Sec.~\ref{sec:Special}.

Under duality transformation \(a_n = \frac{1}{\sqrt{L}}\sum_k b_k e^{i2\pi\alpha k n}\), the Hamiltonian in Eq.~\eqref{eq:hamiltonian} can be transformed into momentum space, yielding
\begin{equation}
	H_k =\sum_k \left( J_+ b^\dag_{k+1}b_k + J_- b^\dag_k b_{k+1} \right) + \sum_k \tilde{V}_k b^\dag_k b_k.
	\label{eq:momentum}
\end{equation}
Here, \( b_k \) (\( b^\dagger_k \)) is the annihilation (creation) operator at quasimomentum \( k \). In this form, the relative phase between the real and imaginary onsite modulations is mapped to generally asymmetric nearest-neighbor couplings \(J_+\) and \(J_-\), whereas the onsite modulation becomes the real function \(\tilde{V}_k\). Explicitly, the system parameters are
\begin{eqnarray}
	J_+ &= & \left(\frac{v_R}{2}-\frac{v_I}{2}\sin\phi\right)+i\frac{v_I}{2}\cos\phi,\nonumber  \\
	J_- &= & \left(\frac{v_R}{2}+\frac{v_I}{2}\sin\phi\right)+i\frac{v_I}{2}\cos\phi,\nonumber  \\
	\tilde{V}_k &=& 2t\cos(2\pi\alpha k).
\end{eqnarray}

For generic \(\phi\), \(|J_+|\neq |J_-|\), and the dual hopping is nonreciprocal. At \(\phi=m\pi\), however, \(J_+=J_-\), so the dual hopping becomes reciprocal even though the Hamiltonian may still be non-Hermitian. In the periodic dual lattice, the asymmetric couplings manifest as point-gap spectral winding, while the corresponding open dual chain would exhibit NHSEs. The interplay of these ingredients--dual nonreciprocity and quasiperiodic modulation--leads to phenomena richer than those produced by either alone, such as asymmetric localized states \cite{NHTQC03} and topological triple phase transitions \cite{TriplePT1,TriplePT2,TriplePT3}. The models in Eqs.~\eqref{eq:hamiltonian} and~\eqref{eq:momentum} are mutually dual and exhibit complementary localization behaviors. Therefore, we expect that the model in Eq.~\eqref{eq:hamiltonian} also hosts rich localization and topological features. Moreover, analyzing the dual model in Eq.~\eqref{eq:momentum} provides a useful approach to understanding the physics of the system in Eq.~\eqref{eq:hamiltonian}.

Before presenting our main results, we introduce the quantities used to characterize the localization properties of the system, namely the IPR and the FD. In numerical calculations, we adopt a rational approximation \(\alpha \approx F_m / F_{m+1}\equiv\alpha_m\) and set the size of the system to \(L = F_{m+1}\), where \(F_m\) denotes the \(m\)th Fibonacci number defined recursively by \(F_{m+1} = F_{m-1} + F_m\) with \(F_1 = F_2 = 1\). This choice allows us to impose periodic boundary conditions and thereby avoid boundary-related artifacts.

\begin{figure*}
	\includegraphics[width=0.95\linewidth]{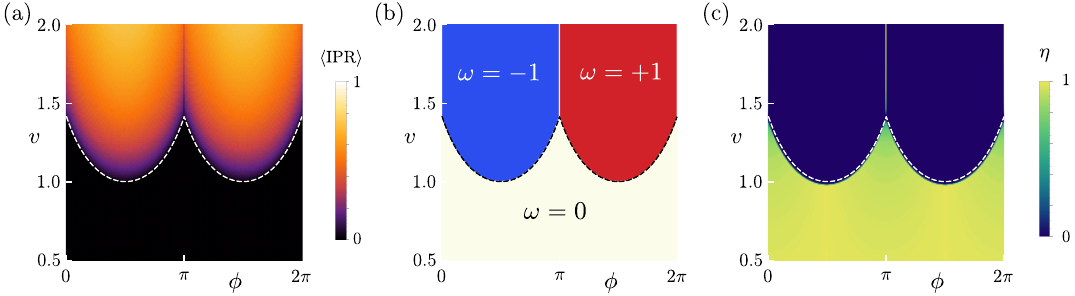}
	\caption{Phase diagrams for topological triple phase transitions. (a) Localization phase diagram: averaged IPR, $\langle \mathrm{IPR} \rangle$, as a function of $v$ and $\phi$. Dark (bright) regions correspond to extended (localized) phases. (b) Topological phase diagram: point-gap winding number $\omega(E_B)$ as a function of $v$ and $\phi$, where the reference energy is chosen as \(E_B=0\). (c) Degeneracy phase diagram: degeneracy rate \(\eta(\epsilon)\) as a function of \(v\) and \(\phi\). In numerical simulations, the system size is set to \(L=F_{15}=610\). For each eigenenergy \(E_i\), we define the nearest spectral distance as \(\delta_i=\min_{j\neq i}|E_i-E_j|\), and regard the state as nearly degenerate when \(\delta_i<\epsilon\), with \(\epsilon=10^{-5}\) (see Appendix \ref{app:degeneracy_threshold} for a justification). Dashed lines denote phase boundaries obtained analytically.}
	\label{phasediagram}
\end{figure*}

The IPR of the \(i\)th right-normalized eigenstate \(|\Psi_i\rangle = \sum_{n=1}^{L} \psi_{i,n} a^{\dag}_{n}|\mathrm{vac}\rangle\) of $H$ is defined as
\begin{equation}
\mathrm{IPR}_i = \sum_{n=1}^{L} |\psi_{i,n}|^4,
\end{equation}
which goes to \(1\) for a localized state and to \(0\) for an extended state when $L\rightarrow\infty$. The FD is further given by
\begin{equation}
\Gamma\equiv\mathrm{FD}_i = -\lim_{L \to \infty} \frac{\ln \mathrm{IPR}_i}{\ln L},
\end{equation}
which approaches \(0\) in localized phases and \(1\) in extended phases for all $i=1,...,L$. We also examine the averaged IPR, defined as
\begin{equation}
\langle \mathrm{IPR} \rangle = \frac{1}{L} \sum_{i=1}^L \mathrm{IPR}_i,
\end{equation}
which approaches \(1\) in localized regimes and \(0\) in extended regimes.
Throughout this work, we set \(t = 1\) and \(v_R = v_I = v\) unless otherwise stated.
	
\section{Phase transitions}\label{sec:Phase}

In this section, we examine localization, topological, and degeneracy-breaking phase transitions driven by the amplitude and phase shift of the onsite potential in our system. All these phase transitions can be understood within the dual momentum-space representation, which features asymmetric couplings. We demonstrate that these three distinct transitions coincide along the same boundary in parameter space.

\subsection{Localization-delocalization transitions}
We first investigate the localization property of the system by computing the averaged IPR, \(\langle \mathrm{IPR} \rangle\), as a function of \(v\) and \(\phi\). The results are shown in Fig.~\ref{phasediagram}(a). Extended phases are characterized by \(\langle \mathrm{IPR} \rangle \simeq 0\), while localized phases exhibit a significantly larger \(\langle \mathrm{IPR} \rangle\). In particular, for \(v\in(1,\sqrt{2})\), varying the relative phase \(\phi\) between the real and imaginary modulations can drive a localization-delocalization transition.

\begin{figure}
	\includegraphics[width=\linewidth]{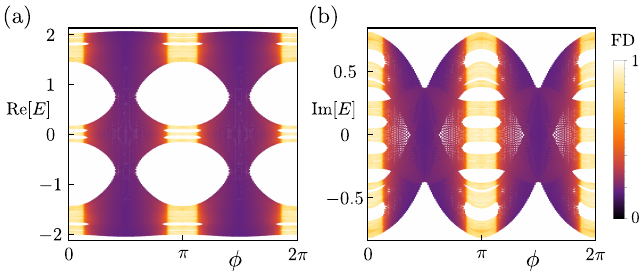}
	\caption{Energy spectrum as a function of the relative phase shift $\phi$, with the fractal dimension $\Gamma$ of the corresponding eigenstates overlaid. (a) Real parts of the spectrum $\mathrm{Re}[E]$. (b) Imaginary parts of the spectrum $\mathrm{Im}[E]$. The amplitude is fixed at $v = 1.2$. In numerical simulations, the system size is set to $L = F_{15} = 610$.}
	\label{spectrum}
\end{figure}

For each given pair of parameters \((\phi, v)\), the eigenstates are either fully extended or localized. As an illustration, we compute the energy spectrum as a function of \(\phi\) for fixed \(v = 1.2\) and mark the fractal dimension \(\Gamma\) of each state. The results are presented in Fig.~\ref{spectrum}, where Figs.~\ref{spectrum}(a) and \ref{spectrum}(b) show the real and imaginary parts of the spectrum, respectively. At \(\phi = 0\), all states are extended. As \(\phi\) increases, the states undergo a localization-delocalization transition beyond a critical point, becoming most strongly localized at \(\phi = \pi/2\). Upon further increasing \(\phi\), the system reverts to being extended. This cycle repeats for every shift of $\phi$ over $2\pi$.

The localization-delocalization transition can be understood from the competition between quasiperiodic-potential-induced Anderson localization and nonreciprocal hoppings in the dual momentum-space Hamiltonian described by Eq.~(\ref{eq:momentum}). We first replace the asymmetric couplings by reciprocal ones with the effective hopping amplitude \(J=\sqrt{|J_+J_-|}\). The dual Hamiltonian then reduces to the standard AAH model with hopping \(J\) and quasiperiodic potential \(2t\cos(2\pi\alpha k)\). This reciprocal model is extended for \(t<J\) and localized for \(t>J\), with the Lyapunov exponent \(\gamma=\ln(t/J)\) in the localized regime. When the asymmetric couplings are restored, the phase boundary is determined by comparing \(\gamma\) with the nonreciprocal asymmetry ratio \(\alpha_J=\ln\sqrt{\max(|J_+|,|J_-|)/\min(|J_+|,|J_-|)}\).

Here, the localized states refer to the quasiperiodic-potential-induced localized states of the reciprocal dual AAH model, rather than the skin-localized states generated by nonreciprocal hopping under open boundary conditions. For such a localized state with localization length \(\xi\), the wave function decays exponentially away from its localization center. When \(L\gg\xi\), changing from open to periodic boundary conditions only affects the exponentially small tails of the localized state. Therefore, in the thermodynamic limit, the localization properties and the corresponding Lyapunov exponent are insensitive to the boundary condition, so the comparison between \(\gamma\) and \(\alpha_J\) remains applicable under periodic boundary conditions.

For \(\gamma>\alpha_J\), the quasiperiodic localization in the dual lattice dominates and the dual states remain localized. For \(\gamma<\alpha_J\), the dual states become extended. Therefore, the transition occurs at \(\gamma=\alpha_J\), which gives
\begin{equation}
	\max(|J_+|,|J_-|)=t.
	\label{eq:dual_transition_condition}
\end{equation}

For \(v_R=v_I=v\), we have \(|J_+|=v\sqrt{(1-\sin\phi)/2}\) and \(|J_-|=v\sqrt{(1+\sin\phi)/2}\), so that \(\max(|J_+|,|J_-|)=v\sqrt{(1+|\sin\phi|)/2}\). Using Eq.~(\ref{eq:dual_transition_condition}), we obtain the phase boundary in the \((v,\phi)\) plane as
\begin{equation}
	v_c = \frac{\sqrt{2}\,t}{\sqrt{1+|\sin\phi|}}.
\end{equation}
Throughout this work we set \(t=1\), and therefore
\begin{equation}
	v_c = \sqrt{\frac{2}{1+|\sin\phi|}}.
	\label{boundary}
\end{equation}
For \(v<v_c\), all eigenstates of the momentum-space Hamiltonian are localized, while for \(v>v_c\) these eigenstates are extended. The real-space Hamiltonian exhibits opposite localization behaviors. The dashed line in Fig.~\ref{phasediagram}(a) shows the phase boundary calculated from Eq.~(\ref{boundary}), which accurately matches the numerical simulation. An independent derivation based on the real-space Lyapunov exponent and Avila's global theory is given in Appendix~\ref{app:avila}. 

In the main text, we primarily adopt periodic boundary conditions. Simulation results for the dual momentum-space Hamiltonian under open boundary conditions are presented in Appendix~\ref{app:skin}, where NHSEs are observed.

\subsection{Topological transitions}

Next, we study the topological properties of the system. Since the energy spectrum of a generic non-Hermitian Hamiltonian is complex, the relevant topological characterization is based on line-gap or point-gap topology. To characterize the spectral topology here, we introduce a magnetic flux \(\Phi\) into the momentum-space Hamiltonian via a twisted boundary condition. The Hamiltonian in Eq.~\eqref{eq:momentum} then becomes
\begin{eqnarray}
	H(\Phi) &= &\sum_{k=1}^{L-1} \left( J_+ b^\dag_{k+1}b_k + J_- b^\dag_k b_{k+1} \right) + \sum_{k=1}^{L} \tilde{V}_k b^\dag_k b_k \nonumber  \\
	&& + J_+ e^{i\Phi} b^\dag_{1}b_L + J_- e^{-i\Phi}b^\dag_L b_{1},  
	\label{flux}
\end{eqnarray}
where the phase \(\Phi\) is applied to the boundary hopping terms. The point-gap winding number is defined with respect to a reference energy \(E_B\) as
\begin{equation}
	\omega(E_B) = \frac{1}{2\pi i}\int_0^{2\pi}\partial_\Phi\ln[\det(H(\Phi)-E_B)] \, d\Phi,
	\label{winding}
\end{equation}
which quantifies how many times the complex spectral trajectory winds around \(E_B\) as \(\Phi\) varies from \(0\) to \(2\pi\). The reference energy \(E_B\) should be chosen inside a point gap of the complex spectrum. In the numerical phase diagram shown in Fig.~\ref{phasediagram}(b), we take \(E_B=0\). As long as the reference energy remains inside the same point gap, the winding number is unchanged.

The winding number is computed as a function of \(v\) and \(\phi\). The results are shown in Fig.~\ref{phasediagram}(b). The parameter space is divided into three regions: two topological regions with \(\omega(E_B=0) = \pm 1\) and a trivial region with \(\omega(E_B=0) = 0\). The topological phase transition from trivial to nontrivial coincides with the transition from extended to localized phases.

\begin{figure}
	\includegraphics[width=\linewidth]{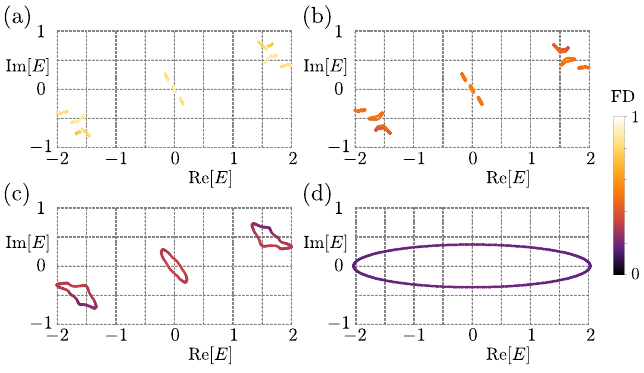}
	\caption{Complex energy spectra for different relative phase shifts $\phi$, with the fractal dimension $\Gamma$ of the corresponding eigenstates overlaid. (a) $\phi = 0$, (b) $\phi = 0.14\pi$, (c) $\phi = 0.2\pi$, and (d) $\phi = 0.5\pi$. The modulation amplitude is fixed at $v = 1.2$. In numerical simulations, the system size is set to $L = F_{15} = 610$.}
	\label{complex}
\end{figure}

The nontrivial topology arises from point-gap spectral windings generated by the asymmetric hopping amplitudes \(J_+\) and \(J_-\) in the dual momentum-space Hamiltonian under periodic boundary conditions. As an illustration, we examine the complex energy spectra at fixed \(v=1.2\) for several values of \(\phi\) across the topological phase transition. The results for \(\phi = 0\), \(0.14\pi\), \( 0.2\pi\), and \(0.5\pi\) are shown in Fig.~\ref{complex}. At \(\phi = 0\), the system is topologically trivial, and the complex spectrum forms clusters without loop structures [Fig.~\ref{complex}(a)]. Near the topological phase transition point (\(\phi = 0.14\pi\)), the spectrum begins to exhibit loop structures, although they remain indistinct [Fig.~\ref{complex}(b)]. Increasing \(\phi\) further into the topological region (\(\phi = 0.2\pi\)), three main loops emerge in the spectrum [Fig.~\ref{complex}(c)], corresponding to the two main band gaps of the underlying AAH model. As \(\phi\) continues to increase, these loops merge into one, as exemplified by the case with \(\phi = 0.5\pi\) in Fig.~\ref{complex}(d).

The topological properties can also be understood from the momentum-space Hamiltonian in Eq.~\eqref{eq:momentum}. For \(v < \sqrt{2/(1+|\sin\phi|)}\), the momentum-space Hamiltonian is in the localized phase: all states are localized, there is no global spectral flow under the boundary twist, and consequently the point-gap winding number \(\omega\) is zero. For \(v > \sqrt{2/(1+|\sin\phi|)}\), the momentum-space Hamiltonian is in the extended phase. The combination of extended states and asymmetric couplings yields a nonzero point-gap winding. Specifically, for $\phi\in(0,\pi)$, we have \(|J_+| < |J_-|\) and the point-gap winding number \(\omega(0)=-1\). For $\phi\in(\pi,2\pi)$, we have \(|J_+| > |J_-|\) and the winding number \(\omega(0)=1\). At \(\phi=m\pi\), the two hopping amplitudes satisfy \(J_+=J_-\), so the nonreciprocity in the dual momentum-space Hamiltonian vanishes. Consequently, the point-gap spectral winding vanishes and the system remains topologically trivial, even though a conventional localization transition can still occur.

\begin{figure*}
	\includegraphics[width=0.95\linewidth]{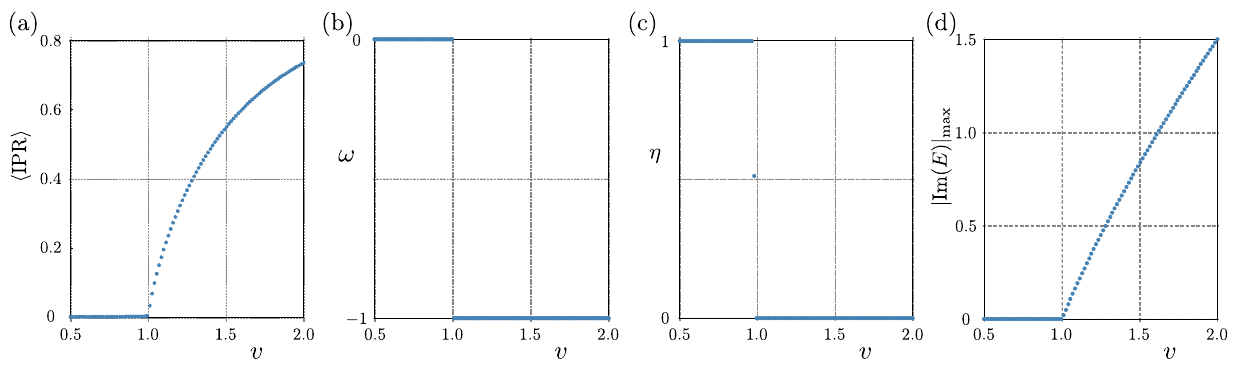}
	\caption{Topological metal-insulator transitions in the ${\cal PT}$-symmetric case with $\phi = \pi/2$. (a) Localization-delocalization transition: averaged IPR, $\langle \mathrm{IPR} \rangle$, vs $v$. (b) Topological transition: point-gap winding number \(\omega(E_B)\) vs \(v\), where the reference energy is chosen as \(E_B=0\). (c) Degeneracy-breaking transition: degeneracy rate \(\eta(\epsilon)\) vs \(v\). (d) ${\cal PT}$ transition: maximum imaginary part of eigenenergies \( |\mathrm{Im}(E)|_{\text{max}} \) vs \(v\). In numerical simulations, the system size is set to \(L=F_{15}=610\). The degeneracy rate is evaluated with the threshold \(\epsilon=10^{-5}\).}
	\label{PTsymmetry}
\end{figure*}

\subsection{Degeneracy-breaking transitions}

Finally, we examine the degeneracy properties of the system. Here, the degeneracy refers to finite-size near-degeneracy rather than an exact symmetry-protected degeneracy. To make the definition of the degeneracy rate precise, for each eigenenergy \(E_i\) we first define its nearest spectral distance as
\begin{equation}
	\delta_i=\min_{j\neq i}|E_i-E_j|.
\end{equation}
An eigenstate is regarded as nearly degenerate when \(\delta_i<\epsilon\), where \(\epsilon\) is a small finite-size threshold. The degeneracy rate is then defined as
\begin{equation}
	\eta(\epsilon)=\frac{1}{L}\sum_{i=1}^L\Theta(\epsilon-\delta_i),
	\label{eq:degeneracy_rate}
\end{equation}
where \(\Theta(x)\) is the Heaviside step function. In the main text, we use \(\epsilon=10^{-5}\) as the numerical threshold for identifying nearly degenerate states. A detailed discussion of how the degeneracy rate varies with the numerical threshold is provided in Appendix~\ref{app:degeneracy_threshold}.

We compute \(\eta(\epsilon)\) as a function of \(v\) and \(\phi\), with the results shown in Fig.~\ref{phasediagram}(c). The parameter space is divided into two regions: one in which most states are nearly twofold degenerate (\(\eta \simeq 1\)), and another in which the states are predominantly nondegenerate (\(\eta \simeq 0\)). The transition between these two regions coincides with the transition from extended to localized phases, as well as with the topological transition from trivial to nontrivial phases, with slight deviations due to finite-size effects.

The degeneracy-breaking transition can be understood by analyzing the momentum-space Hamiltonian in Eq.~(\ref{eq:momentum}). In the limit $v/t \rightarrow 0$, all states are localized in momentum space. The states $|k_0\rangle$ and $|-k_0\rangle$ are exactly degenerate, forming a degenerate subspace. Extending to the more general localized phase of the momentum-space Hamiltonian, the corresponding localized states are centered at $k_0$ and $-k_0$ and are therefore separated by a distance $2|k_0|$ in momentum space. Their wave-function overlap, denoted by $\mathcal{O}_{k_0,-k_0}$, decays exponentially with this separation as $\mathcal{O}_{k_0,-k_0}\sim e^{-2|k_0|/\xi}$, where $\xi$ is the localization length. The coupling-induced energy splitting is proportional to this overlap, $\Delta\propto |v|\,\mathcal{O}_{k_0,-k_0}$. Therefore, when $2|k_0|$ is sufficiently large, $\mathcal{O}_{k_0,-k_0}\rightarrow0$ and consequently $\Delta\rightarrow0$. Hence, in the localized phase of the momentum-space Hamiltonian, most states are nearly doubly degenerate.

We next consider the opposite limit \(t/v\rightarrow 0\), where the dual momentum-space Hamiltonian reduces to a uniform nonreciprocal chain. The lifting of the twofold degeneracy should not be attributed simply to time-reversal-symmetry breaking. The relevant symmetry is instead the reciprocal \(n\leftrightarrow -n\) spectral symmetry. In the uniform limit, the dispersion relation of the dual chain is
\begin{equation}
	E(n)=J_+e^{-in}+J_-e^{in}.
	\label{eq:dual_dispersion}
\end{equation}
For reciprocal hopping \(J_+=J_-\), one has \(E(n)=E(-n)\), so the counterpropagating modes \(n\) and \(-n\) are degenerate. In contrast, when \(J_+\neq J_-\), this \(n\leftrightarrow -n\) spectral symmetry is generally lost, and the twofold degeneracy between the counterpropagating modes is lifted. This remains true even when \(J_+\) and \(J_-\) are both real, where complex-conjugation symmetry may still be present but does not enforce \(E(n)=E(-n)\).

Because the momentum-space and real-space Hamiltonians exhibit complementary localization behaviors, in the extended phase of the real-space Hamiltonian, most states are nearly doubly degenerate, whereas in the localized phase, the states are predominantly non-degenerate. Therefore, the degeneracy-breaking transition is tied to the same boundary as the Anderson localization and point-gap topological transitions: below the boundary, the dual momentum-space states are localized and the effective \(n\leftrightarrow -n\) pairing remains nearly degenerate, while above the boundary, the dual states become extended and the nonreciprocal hopping resolves the counterpropagating pairs into non-degenerate eigenstates.

\section{Phase transitions at special phase shifts}\label{sec:Special}

We next examine the phase transitions at specific high symmetry points in the parameter space $\phi$, i.e., $\phi = m\pi + \pi/2$ and $\phi = m\pi$. At $\phi = m\pi + \pi/2$, the system acquires the ${\cal PT}$ symmetry, which introduces an additional real-complex spectral transition alongside the previously discussed triple phase transitions. In contrast, the dual hopping becomes reciprocal at \(\phi=m\pi\), and only the localization-delocalization transition remains.
\vspace{-0.6em}
\subsection{Phase transitions at \texorpdfstring{${\phi=\pi/2}$}{phi=pi/2}}

We use $\phi = \pi/2$ as an example to illustrate the phase transitions at phase shift $\phi = m\pi + \pi/2$. Here, the quasiperiodic onsite potential becomes
\begin{equation}
	V_n = v_R \cos(2\pi \alpha n) - i \, v_I \sin( 2\pi \alpha n ),
\end{equation}
which satisfies $V_{-n} = V_n^*$. The Hamiltonian $H$ then possesses ${\cal PT}$ symmetry.

${\cal PT}$-symmetric NHQCs could support a triple phase transition, encompassing a localization-delocalization transition, a topological transition, and a ${\cal PT}$-symmetry breaking transition (or a real-complex spectral transition) simultaneously. Here, we compute the averaged IPR $\langle \mathrm{IPR} \rangle$, the point-gap winding number \(\omega(0)\), the degeneracy rate \(\eta(\epsilon)\), and the maximum imaginary part of the eigenenergies $|\mathrm{Im}(E)|_{\text{max}}$ as functions of $v$. The results are presented in Fig.~\ref{PTsymmetry}. We observe that all phase transitions occur at $v = 1$, as described by Eq.~(\ref{boundary}). Thus, for ${\cal PT}$-symmetric NHQCs, we find a quartet phase transition, whereas in the general case without ${\cal PT}$ symmetry, only a triple phase transition is present.

\subsection{Phase transitions at \texorpdfstring{${\phi=0}$}{phi=0}}

\begin{figure}
	\includegraphics[width=0.95\linewidth]{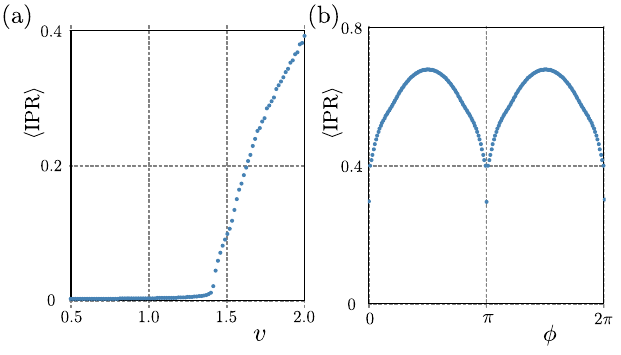}
	\caption{Localization-delocalization transition for $\phi=0$. (a) Averaged IPR, $\langle \mathrm{IPR} \rangle$, as a function of $v$ for $\phi=0$. (b) $\langle \mathrm{IPR} \rangle$ as a function of $\phi$ for $v=1.8$.  In numerical simulations, the system size is set to $L = F_{15} = 610$.}
	\label{general}
\end{figure}

We use \(\phi=0\) as an example to illustrate the phase transitions at the phase shift \(\phi=m\pi\). At \(\phi=0\), we have \(J_+=J_-\), so the dual momentum-space Hamiltonian is reciprocal and does not exhibit point-gap spectral winding under periodic boundary conditions. Consequently, no topological phase transitions occur, as evidenced by the vanishing winding number for all \(v\) at \(\phi=m\pi\) in Fig.~\ref{phasediagram}(b). The condition \(J_+=J_-\) also restores the reciprocal \(n\leftrightarrow -n\) spectral symmetry, so that the degeneracy-breaking transition is absent, consistent with \(\eta \simeq 1\) for all \(v\) at \(\phi=m\pi\) in Fig.~\ref{phasediagram}(c). Nevertheless, the localization-delocalization transition remains. To examine this, we compute the averaged IPR, \(\langle \mathrm{IPR} \rangle\), as a function of \(v\) at \(\phi=0\), with the results shown in Fig.~\ref{general}(a). A clear change is observed around the phase transition point \(v=\sqrt{2}\), consistent with the phase boundary described by Eq.~(\ref{boundary}). This suggests that two types of localization-delocalization transitions may exist in general NHQCs: one accompanied by point-gap spectral topology and degeneracy-breaking, and the other analogous to that in Hermitian systems, where only localization changes.

To further illustrate the localization properties of the system, the averaged IPR, \(\langle \mathrm{IPR} \rangle\), is computed as a function of \(\phi\) at a fixed \(v = 1.8\), as shown in Fig.~\ref{general}(b). For all \(\phi\), the system remains in the localized phase, as indicated by the finite \(\langle \mathrm{IPR} \rangle\) values. However, the value of \(\langle \mathrm{IPR} \rangle\) at \(\phi = m\pi\) is significantly lower than at other values of \(\phi\), implying a relatively longer localization length in this reciprocal dual-hopping case. This behavior is consistent with the absence of point-gap spectral winding and degeneracy-breaking at \(\phi=m\pi\).

In summary, the essential phase-transition properties of our system are presented in Table.~\ref{table:summary}.

\begin{table*}[!t]
\caption{Summary of the main phase-transition properties. The localization transition is determined by the dual-space competition between quasiperiodic localization and hopping asymmetry. The topological transition is characterized by the point-gap winding, while the degeneracy-breaking transition originates from the loss of reciprocal \(n\leftrightarrow -n\) spectral symmetry.}
\label{table:summary}
\begin{ruledtabular}
\begin{tabular*}{\textwidth}{@{\extracolsep{\fill}}c c c c c}
Phase regime & Dual momentum-space behavior & Real-space behavior & Point-gap winding & Degeneracy\\
\hline
\(v<v_c\) & Localized & Extended & \(\omega=0\) & Near-degenerate\\
\(v>v_c\) & Extended & Localized & \(\omega=\pm1\) & Non-degenerate\\
\(\phi=m\pi\) & Reciprocal dual hopping & Localization transition only & $\omega=0$ & Near-degenerate\\
\(\phi=m\pi+\pi/2\) & \({\cal PT}\)-symmetric & Quartet transition & $\omega=0\,(v<v_c),\pm1\,(v>v_c)$ & Degeneracy breaking
\end{tabular*}
\end{ruledtabular}
\end{table*}

\vspace{-0.5em}
\section{Conclusion}\label{sec:Sum}

We proposed a non-Hermitian quasiperiodic model that generally does not preserve ${\cal PT}$ symmetry and demonstrated that the system supports generic triple phase transitions that involve localization, topology and degeneracy. The topological transition is associated with the point-gap spectral winding in the dual momentum-space Hamiltonian, while the degeneracy-breaking transition follows from the loss of the reciprocal \(n\leftrightarrow -n\) spectral symmetry. The system also exhibits a localization transition analogous to the Hermitian case, namely, it occurs without involving a topological or a  degeneracy-breaking transition. In most regions of the parameter space, the localization, topological, and degeneracy-breaking transitions take place simultaneously, whereas at special phase shifts, such as $\phi=m\pi$, only localization transitions are present. Our findings extend the study of localization and topological transitions in NHQCs beyond the ${\cal PT}$-symmetric framework and motivate further experimental investigations of phase transitions in quasicrystal metamaterials, where the relative phase shift between the real and imaginary quasiperiodic modulations may serve as a flexible control parameter for non-Hermitian phase transitions.

Our model may be emulated in photonic quantum-walk platforms, where
quasiperiodic phase modulations, controlled optical losses, and effective
nonreciprocal couplings can be engineered independently
~\cite{WeidemannNature2022,LinPRL2022}. Time-multiplexed photonic quantum
walks based on coupled optical-fiber loops have already been employed to
realize non-Hermitian Floquet quasicrystals and to experimentally probe
localization transitions, point-gap spectral topology, non-Hermitian skin
effects, and multiple coincident phase transitions. In these platforms,
electro-optic phase modulators can generate quasiperiodic phase potentials,
whereas acousto-optic amplitude modulators introduce spatially controlled
gain and loss. These established techniques suggest that the relative phase
\(\phi\) in the present model could be implemented by shifting the
loss-modulation pattern with respect to the quasiperiodic phase-modulation
pattern. The localization properties may be detected from the spatial
spreading of the photon distribution and the corresponding dynamical
inverse participation ratio, while non-Hermitian boundary accumulation may
be observed by imposing open boundary conditions. Although the existing
Floquet quantum-walk realizations are not term-by-term identical to the
static Hamiltonian considered here, they provide a feasible experimental
route for probing the predicted phase transitions.

Several recent works have also studied non-Hermitian quasiperiodic models with asymmetric hopping, complex onsite potentials, and relative phases between real and imaginary modulations~\cite{AcharyaDatta2024,CaiJiang2022}. These studies provide important background for understanding localization and spectral properties in non-Hermitian quasicrystals. The focus of the present work is complementary. Instead of starting from a system with explicitly nonreciprocal real-space hopping, we consider a model with reciprocal nearest-neighbor hopping but with a complex quasiperiodic onsite potential whose real and imaginary parts are shifted by a tunable relative phase. This relative phase reorganizes the Fourier components of the onsite potential and, after duality transformation, appears as asymmetric hopping amplitudes in momentum space.

\begin{acknowledgments}
W.Z. acknowledges support by the National Natural Science Foundation of China (Grant No.~12404499).
L.Z. acknowledges support by the National Natural Science Foundation of China (Grant No.~12275260),
the Shandong Provincial Natural Science Foundation (Grant No.~ZR2026QB21),
and the Fundamental Research Funds for the Central Universities (Grant No. 202364008).
J.G. acknowledges support by the National Research Foundation, Singapore, through the National Quantum Office, hosted in A*STAR, under its Centre for Quantum Technologies Funding Initiative (S24Q2d0009).
\end{acknowledgments}

\appendix
\section{Real-space Lyapunov exponent from Avila's global theory}
\label{app:avila}

In this Appendix, we provide an alternative derivation of the localization boundary under more general conditions from the real-space quasiperiodic transfer matrix using Avila's global theory \cite{Avila2015}. The real-space eigenvalue equation is
\begin{equation}
	t\psi_{n+1}+t\psi_{n-1}+V(\theta_n)\psi_n=E\psi_n,\quad \theta_n=2\pi\alpha n.
\end{equation}
The onsite potential can be written as
\begin{equation}
	V(\theta)=u_+e^{i\theta}+u_-e^{-i\theta},
\end{equation}
where
\begin{equation}
u_+=\frac{v_R}{2}+i\frac{v_I}{2}e^{i\phi},\qquad
u_-=\frac{v_R}{2}+i\frac{v_I}{2}e^{-i\phi}.
\end{equation}
The corresponding transfer matrix is
\begin{equation}
	A_E(\theta)=
	\begin{pmatrix}
	\dfrac{E-u_+e^{i\theta}-u_-e^{-i\theta}}{t} & -1\\
	1 & 0
	\end{pmatrix}.
\end{equation}
We introduce the complexified quasiperiodic phase shift as \(\theta\rightarrow\theta+i\epsilon\). For \(\epsilon\rightarrow+\infty\), the term proportional to \(u_-\) dominates, and the Lyapunov exponent becomes
\begin{equation}
L(E;\epsilon)=\lim_{\epsilon\rightarrow\infty}\ln{||A_E(\theta+i\epsilon)||}\simeq\epsilon+\ln\frac{|u_-|}{t}+o(1).
\end{equation}
For \(\epsilon\rightarrow-\infty\), the term proportional to \(u_+\) dominates, and the Lyapunov exponent becomes
\begin{equation}
	L(E;\epsilon)=\lim_{\epsilon\rightarrow-\infty}\ln{||A_E(\theta+i\epsilon)||}\simeq-\epsilon+\ln\frac{|u_+|}{t}+o(1).
\end{equation}
By Avila's global theory, \(L(E;\epsilon)\) is a convex piecewise-linear function of \(\epsilon\) with quantized slopes $\pm1$. Thus, in the thermodynamic limit, the Lyapunov exponent satisfies
\begin{equation}
	L(E)=\max\left\{0,\ln\frac{|u_+|}{t},\ln\frac{|u_-|}{t}\right\}.
\end{equation}
Since \(u_\pm\) have the same algebraic form as the \(J_\pm\) in the main text, the localization transition condition under periodic boundary conditions is
\begin{equation}
	\max(|J_+|,|J_-|)=t.
\end{equation}
For \(v_R=v_I=v\), this gives
\begin{equation}
	v_c=\frac{\sqrt{2}\,t}{\sqrt{1+|\sin\phi|}},
\end{equation}
or \(v_c=\sqrt{2/(1+|\sin\phi|)}\) for \(t=1\), in agreement with Eq.~(\ref{boundary}) in the main text.

\setlength{\textfloatsep}{8pt}
\section{Non-Hermitian skin effects in the dual momentum-space lattice}
\label{app:skin}
\begin{figure}
	\includegraphics[width=\linewidth]{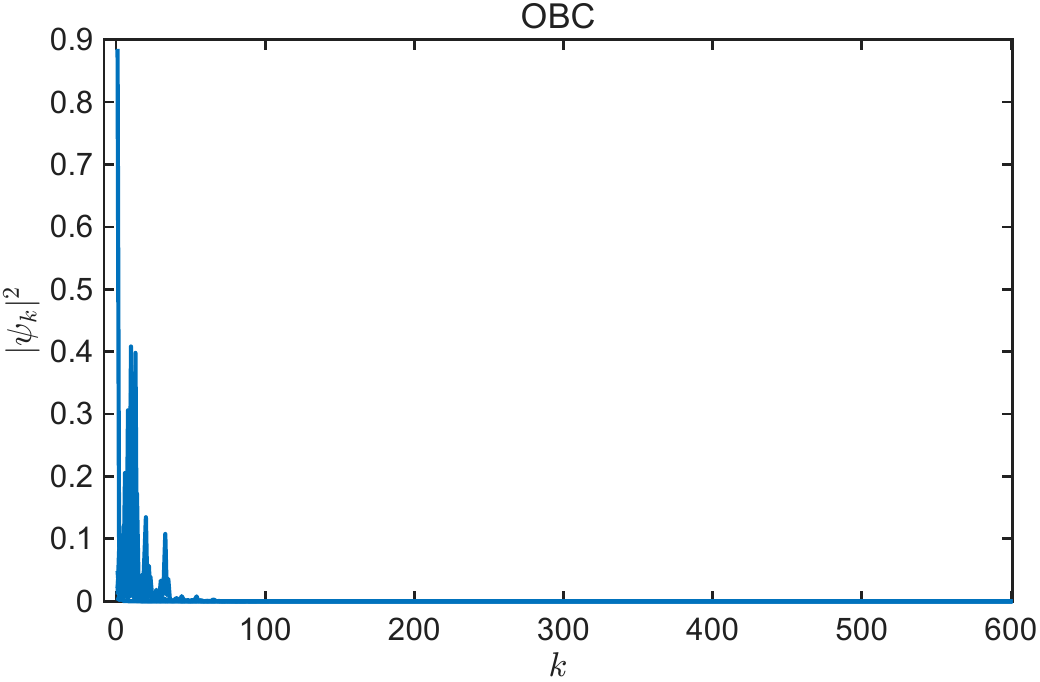}
	\caption{Non-Hermitian skin effects under open boundary conditions. 
	The system parameters are set to \(t=1\), \(v_R=v_I=v=1.2\), and
	\(\phi=0.2\pi\). The probability density
	\(|\psi_k|^2\) accumulates at one boundary of the open dual chain,
	indicating the presence of skin accumulation.}
	\label{OBC}
\end{figure}

The asymmetric hopping in the dual momentum-space Hamiltonian has distinct
manifestations under periodic and open boundary conditions. Under periodic
boundary conditions, the asymmetric couplings generate point-gap spectral
winding, which is used to characterize the topological transition in the main
text. Under open boundary conditions, the same nonreciprocal hopping leads to
the non-Hermitian skin effect, where bulk eigenstates accumulate at one
boundary.

To visualize this skin accumulation, we consider the dual momentum-space Hamiltonian in
Eq.~\eqref{eq:momentum} under open boundary conditions by removing the boundary
hopping terms. For the representative parameters
\(t=1\), \(v_R=v_I=v=1.2\), and \(\phi=0.2\pi\), the dual hopping amplitudes satisfy
\(|J_+|\neq |J_-|\). The probability distributions of representative right
eigenstates are shown in Fig.~\ref{OBC}. All eigenstates here accumulate
at the left boundary, demonstrating non-Hermitian skin effects induced
by asymmetric dual hoppings.

\newpage
\section{Threshold dependence of the degeneracy rate}
\label{app:degeneracy_threshold}

The definition of finite-size degeneracy rate \(\eta(\epsilon)\) involves a numerical threshold \(\epsilon\), which is used to determine whether two eigenenergies are nearly degenerate. To verify that the degeneracy-breaking phase boundary identified in the main text does not depend sensitively on this threshold, we repeat the calculations here with \(\epsilon=10^{-4}\), \(10^{-5}\), and \(10^{-6}\).

\begin{figure}[H]
	\includegraphics[width=\linewidth]{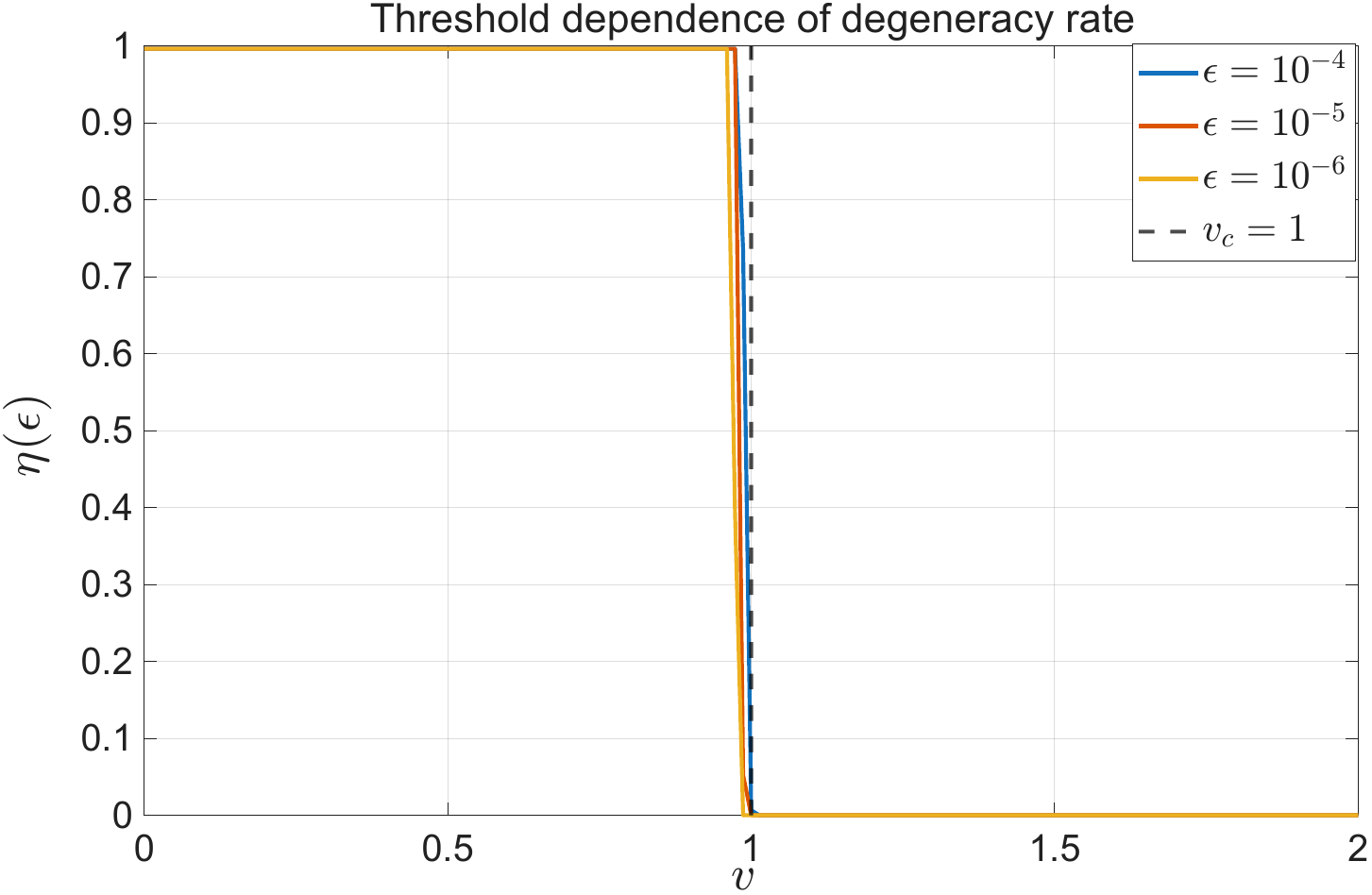}
	\caption{Threshold dependence of the degeneracy rate \(\eta(\epsilon)\). The degeneracy rate is calculated using three different thresholds, \(\epsilon=10^{-4}\), \(10^{-5}\), and \(10^{-6}\), for identifying nearly degenerate states through the condition \(\delta_i<\epsilon\), where \(\delta_i=\min_{j\neq i}|E_i-E_j|\). Although the absolute value of \(\eta(\epsilon)\) varies slightly with the threshold, the transition point and the overall phase boundary remain essentially unchanged.}
	\label{degeneracy_threshold}
\end{figure}

As shown in Fig.~\ref{degeneracy_threshold}, changing \(\epsilon\) modifies the absolute value of \(\eta(\epsilon)\) because a larger threshold counts more eigenenergy pairs as nearly degenerate. Nevertheless, the position of the sharp change in \(\eta(\epsilon)\) and the corresponding phase boundary remain nearly unchanged for all three threshold values. Therefore, the degeneracy-breaking transition discussed in the main text is robust against a reasonable variation of the finite-size threshold.

\color{black}
\bibliography{references}
	
\end{document}